\documentclass[12pt]{article}
\usepackage{bm,amsmath,amssymb,graphicx,mathrsfs,slashed}

\begin{document}
\pagenumbering{arabic}
\begin{titlepage}

\title{Conformal theory of everything}

\author{F. F. Faria$\,^{*}$ \\
Centro de Ci\^encias da Natureza, \\
Universidade Estadual do Piau\'i, \\ 
64002-150 Teresina, PI, Brazil}

\date{}
\maketitle

\begin{abstract}
By using conformal symmetry we unify the standard model of particle 
physics with gravity in a consistent quantum field theory which 
describes all the fundamental particles and forces of nature. 
\end{abstract}

\thispagestyle{empty}
\vfill
\noindent PACS numbers: 104.60.-m, 98.80.-k, 04.50.+h \par
\bigskip
\noindent * felfrafar@hotmail.com \par
\end{titlepage}
\newpage


\section{Introduction}
\label{sec1}


It is well know that the standard model (SM) of particle physics is consistent 
with the experiments performed so far on particle accelerators such as the 
large hadron collider (LHC). However, the theory presents some problems 
such as the hierarchy and the Landau pole problems. Several modifications 
of the SM at scales between the electroweak and Planck scales, such as 
GUT \cite{Georgi1,Georgi2,Fritz,Gursey} or SUSY \cite{Miyazawa,Yu,Gervais,
Volkov,Wess,Salam,Fayet,Nilles,Haber}, have been proposed to solve such 
problems. However, it is likely that there is no new physics beyond the SM 
all the way up to the Planck scale \cite{Frog}. This leads us to suppose 
that the SM is a low energy limit of a fundamental theory defined at the 
Planck scale.  Since gravitational effects are expected to be important 
around the Planck scale, it is natural to conjecture that this fundamental 
theory includes quantum gravity.

One of the most straightforward way to extend and unify physical theories 
is to change their symmetries. A strong candidate to be 
incorporated in the unification of the SM with gravity is the (local) 
conformal symmetry, which perform a multiplicative rescaling of all fields 
according to
\begin{equation}
\tilde{\Phi} = \Omega(x)^{-\Delta_{\Phi}}\Phi,
\label{1}
\end{equation}
where $\Omega(x)$  is an arbitrary function of the spacetime coordinates, 
and $\Delta_{\Phi}$ is the scaling dimension of the field $\Phi$, whose 
values are $-2$ for the metric field, $0$ for gauge bosons, $1$ for 
scalar fields and $3/2$ for fermions.

The consideration of the conformal symmetry as one of the 
fundamental symmetries of physics is justified for several reasons. 
First of all, the classical SM action and the power spectrum of the cosmic 
fluctuations are approximately conformal invariant. Second, the conformal 
symmetry might play an important role in the solution of both the hierarchy 
and the Landau pole problems \cite{Gorsky}. In addition, the black 
hole complementarity principle can be explained with the use of the 
conformal symmetry \cite{Hooft1}. Last but not least, quantum field 
theories in curved spacetime (for a nice review, see \cite{Buch1}) may 
become asymptotically conformally invariant in the limit of a strong 
gravitational field \cite{Buch2}.

The aim of this work is to consistently unify the SM with gravity through 
the use of conformal symmetry. In Section \ref{sec2} we describe a consistent 
conformal theory of quantum gravity, called massive conformal gravity (MCG). 
In Section \ref{sec6} we develop an extended SM conformally coupled with 
gravity. In Section \ref{sec11} we combine MCG with the extended SM to form 
what we call conformal theory of everything (CTOE). Finally, in Section 
\ref{sec12} we present our conclusions.


\section{Gravity}
\label{sec2}


The conformally invariant theory of gravity that we consider here is MCG, whose 
action is given by\footnote{This action is obtained from the 
action of Ref. \cite{Faria1} by the reparametrizations  $\beta\lambda^{-2}/2kc 
\rightarrow 1$ and $\alpha/kc \rightarrow 1/\alpha^2$. Additionally, we included in 
(\ref{2}) the Gauss-Bonnet term  $\int{d^{4}x} \, \sqrt{-g} E$ in order 
to provide renormalizability.} \cite{Faria1}
\begin{equation}
S_{\textrm{MCG}} = \int{d^{4}x} \, \sqrt{-g}\left[ \varphi^{2}R 
+ 6\partial^{\mu}\varphi\partial_{\mu}\varphi - \frac{1}{2\alpha^2}
C^2  - \eta E \right], 
\label{2}
\end{equation}
where $\varphi$ is a scalar field called dilaton, $\alpha$ and $\eta$ are 
dimensionless coupling constants, 
\begin{equation}
C^2 = C^{\alpha\beta\mu\nu}C_{\alpha\beta\mu\nu} = E 
+ 2W
\label{3}
\end{equation}
is the Weyl tensor squared,
$E = R^{\alpha\beta\mu\nu}R_{\alpha\beta\mu\nu}- 4R^{\mu\nu}R_{\mu\nu} + R^2 $ 
is the Euler density, $W = R^{\mu\nu}R_{\mu\nu} 
-(1/3)R^{2}$, $R^{\alpha}\,\!\!_{\mu\beta\nu}$ is the 
Riemann tensor, $R_{\mu\nu} = R^{\alpha}\,\!\!_{\mu\alpha\nu}$ is the 
Ricci tensor, and $R = g^{\mu\nu}R_{\mu\nu}$ is the scalar curvature. 
Considering that $\sqrt{-g}E$ is a total derivative, and integrating by parts, 
we can write (\ref{2}) as
\begin{equation}
S_{\textrm{MCG}} = \int{d^{4}x} \, \sqrt{-g}\left[ \varphi^{2}R 
- 6\varphi\Box\varphi - \frac{1}{\alpha^2} W \right], 
\label{4}
\end{equation}
where $\Box = g^{\mu\nu}\nabla_{\mu}\nabla_{\nu}$ is the generally 
covariant d'Alembertian.

The variation of (\ref{4}) with respect to $g^{\mu\nu}$ and $\varphi$ gives
the field equations
\begin{equation}
\varphi^{2}G_{\mu\nu} +  6 \partial_{\mu}\varphi\partial_{\nu}\varphi 
- 3g_{\mu\nu}\partial^{\rho}\varphi\partial_{\rho}\varphi + g_{\mu\nu} 
 \Box \varphi^{2} 
- \nabla_{\mu}\nabla_{\nu} \varphi^{2}  - \alpha^{-2} B_{\mu\nu} = 0,
\label{5}
\end{equation}
\begin{equation}
\varphi R - 6\Box\varphi  = 0,
\label{6}
\end{equation}
where
\begin{equation}
B_{\mu\nu} = \nabla^{\alpha}\nabla^{\beta}C_{\mu\alpha\nu\beta} 
-\frac{1}{2} R^{\alpha\beta}C_{\mu\alpha\nu\beta} 
\label{7}
\end{equation}
is the Bach tensor, and
\begin{equation}
G_{\mu\nu} = R_{\mu\nu} - \frac{1}{2}g_{\mu\nu}R
\label{8}
\end{equation}
is the Einstein tensor. 

In addition to the conformally invariant field equations (\ref{5}) 
and (\ref{6}), MCG also has conformally invariant line element  
\begin{equation}
ds^2 = (\varphi/\varphi_c)^2g_{\mu\nu}dx^\mu dx^\nu,
\label{9}
\end{equation}
and geodesic equation 
\begin{equation}
\frac{d^{2}x^{\lambda}}{d\tau^2} + \Gamma^{\lambda}\,\!\!_{\mu\nu}
\frac{dx^{\mu}}{d\tau}\frac{dx^{\nu}}{d\tau} +\frac{1}{\varphi}
\frac{\partial\varphi}{\partial x^{\rho}} \left( g^{\lambda\rho} + 
\frac{dx^{\lambda}}{d\tau}\frac{dx^{\rho}}{d\tau}\right) = 0,
\label{10}
\end{equation}
where $\varphi_c$ is a constant scalar field and
\begin{equation}
\Gamma^{\lambda}\,\!\!_{\mu\nu} = \frac{1}{2}g^{\lambda\rho}\left( 
\partial_{\mu}g_{\nu\rho} + \partial_{\nu}g_{\mu\rho} 
- \partial_{\rho}g_{\mu\nu} \right)
\label{11}
\end{equation}
is the Levi-Civita connection. Although the theory has not yet been fully 
tested with classical tests\footnote{It has been shown so far that MCG is 
consistent with solar system observations \cite{Faria2}, has no vDVZ 
discontinuity \cite{Faria3}, can reproduce the orbit of binaries by the 
emission of gravitational waves \cite{Faria4} and describes the late 
universe without the cosmological constant problem \cite{Faria5}.}, only 
its quantum behavior, which will be addressed in more detail next, is 
relevant to this work.


\subsection{Renormalizability}
\label{sec3}


By using the flat background field expansions
\begin{equation}
g_{\mu\nu} = \eta_{\mu\nu} + \alpha h_{\mu\nu}, \ \ \ \ \ 
\varphi = \varphi_{c} + \sigma,
\label{12}
\end{equation}
and keeping only the terms of second order in the quantum fields $h^{\mu\nu}$ 
and $\sigma$, we find that (\ref{4}) reduces to the linearized action
\begin{equation}
\bar{S}_{\textrm{MCG}} =  \int{d^{4}x} 
\Bigg[ m^2\bar{\mathcal{L}}_{EH} + 
2m\sigma\bar{R} - 6 \sigma\bar{\Box}\sigma 
- \left(\bar{R}^{\mu\nu}\bar{R}_{\mu\nu} - \frac{1}{3}
\bar{R}^{2} \right)  \Bigg],
\label{13}
\end{equation}
where $m=\alpha\varphi_c$ is the graviton mass,
\begin{equation}
\bar{\mathcal{L}}_{EH} = - \frac{1}{4} \Big( \partial^{\rho}
h^{\mu\nu}\partial_{\rho}h_{\mu\nu} - 2\partial^{\mu}h^{\nu\rho}
\partial_{\rho}h_{\mu\nu}+ 2\partial^{\mu}h_{\mu\nu}\partial^{\nu}h 
- \partial^{\mu}h\partial_{\mu}h  \Big)
\label{14}
\end{equation}
is the linearized Einstein-Hilbert Lagrangian density,
\begin{equation}
\bar{R}_{\mu\nu} = \frac{1}{2} \left( \partial_{\mu}\partial^{\rho}
h_{\rho\nu} + \partial_{\nu}\partial^{\rho}h_{\rho\mu} 
- \bar{\Box} h_{\mu\nu} 
- \partial_{\mu}\partial_{\nu}h  \right)
\label{15}
\end{equation}
is the linearized Ricci tensor, and
\begin{equation}
\bar{R} =  \partial^{\mu}\partial^{\nu}h_{\mu\nu} 
- \bar{\Box} h
\label{16}
\end{equation} 
is the linearized scalar curvature, with $\bar{\Box}= 
\eta^{\mu\nu}\partial_{\mu}\partial_{\mu}$ and $h = \eta^{\mu\nu}h_{\mu\nu}$. 

The linearized action (\ref{13}) is invariant under the coordinate gauge 
transformation
\begin{equation}
h_{\mu\nu} \rightarrow h_{\mu\nu} + \partial_{\mu}\chi_{\nu} + 
\partial_{\nu}\chi_{\mu},
\label{17}
\end{equation}
where $\chi^{\mu}$ is an arbitrary spacetime dependent vector 
field, and under the conformal gauge transformations
\begin{equation}
h_{\mu\nu} \rightarrow h_{\mu\nu} + \eta_{\mu\nu}\Lambda,
\ \ \ \ \
\sigma \rightarrow \sigma - \frac{1}{2}\Lambda,
\label{18}
\end{equation}
where $\Lambda$ is an arbitrary spacetime dependent scalar field. As usual, 
these gauge symmetries require the addition of gauge fixing and Faddeev–Popov 
ghost terms. 

In order to no part of the graviton propagator to fall off slower than $p^{-4}$ 
for large momenta, we choose the gauge fixing terms 
\begin{equation}
\bar{S}_{GF1} = - \frac{1}{2\xi_{1}}\int{d^{4}x}\,\left(m^2 - \bar{\Box}\right)
\left( \partial^{\mu}h_{\mu\nu} - \frac{1}{2}\partial_{\nu}h \right)^{2},  
\label{19}
\end{equation}
\begin{equation}
\bar{S}_{GF2} =   \frac{1}{6\xi_{2}}\int{d^{4}x}
\left( \bar{R} - 6\xi_{2}m\sigma\right)^{2},  
\label{20}
\end{equation}
and its corresponding Faddeev–Popov ghost terms
\begin{equation}
\bar{S}_{FP1} = \int{d^{4}x} \,
\tilde{c}^{\mu}\left( \bar{\Box} - m^2\right)\bar{\Box} c_{\mu},  
\label{21}
\end{equation}
\begin{equation}
\bar{S}_{FP2} =  2\int{d^{4}x} \, \tilde{c}\left( \bar{\Box} - \xi_{2}m^2\right)c,  
\label{22}
\end{equation}
where $\xi_{1}$ and $\xi_{2}$ are coordinate and conformal gauge fixing 
parameters, ($\tilde{c}^{\mu}$)$c^{\mu}$ is a vector (anti-ghost)ghost field, 
and ($\tilde{c}$)$c$ is a scalar (anti-ghost)ghost field.

Integrating by parts, and performing a long but straightforward calculation, 
we can write $\bar{S} = \bar{S}_{\textrm{MCG}} + \bar{S}_{\textrm{GF1}} 
+ \bar{S}_{\textrm{G2}} + \bar{S}_{\textrm{FP1}} + \bar{S}_{\textrm{FP2}}$ 
in the form
\begin{eqnarray}
\bar{S} &=&  - \int{d^{4}x} 
\bigg\{ \frac{1}{4}h^{\mu\nu}\bigg[\left( \bar{\Box} - m^2 \right)\bar{\Box} 
P^{(2)}_{\mu\nu,\alpha\beta} - \frac{2}{\xi_{2}}\left(\bar{\Box} - \xi_{2}m^2 
\right)\bar{\Box} P^{(0-s)}_{\mu\nu,\alpha\beta} \nonumber \\ &&
+ \frac{1}{2\xi_1} \left(\bar{\Box} - m^2 \right) 
\bar{\Box}\bigg(2P^{(1)}_{\mu\nu,\alpha\beta} +3P^{(0-s)}_{\mu\nu,\alpha\beta}
- \sqrt{3}\left(P^{(0-sw)}_{\mu\nu,\alpha\beta} 
- P^{(0-ws)}_{\mu\nu,\alpha\beta}\right) \nonumber \\ &&
 + P^{(0-w)}_{\mu\nu,\alpha\beta} 
\bigg) \bigg]h^{\alpha\beta} + 6 \sigma \left(\bar{\Box} -\xi_{2}m^2\right)\sigma 
- \tilde{c}^{\mu}\left( \bar{\Box} - m^2\right)\bar{\Box} c_{\mu} \nonumber \\ &&
 - 2\tilde{c}\left(\bar{\Box} - \xi_{2}m^2\right)c \bigg\}, 
\label{23}
\end{eqnarray}
where
\begin{equation}
P^{(2)}_{\mu\nu,\alpha\beta} = \frac{1}{2}\left(\theta_{\mu\alpha}
\theta_{\nu\beta} +\theta_{\mu\beta}\theta_{\nu\alpha}\right) 
- \frac{1}{3}\theta_{\mu\nu}\theta_{\alpha\beta},
\label{24}
\end{equation}
\begin{equation}
P^{(1)}_{\mu\nu,\alpha\beta} = \frac{1}{2}\left(\theta_{\mu\alpha}
\omega_{\nu\beta} +\theta_{\mu\beta}\omega_{\nu\alpha} + \theta_{\nu\alpha}
\omega_{\mu\beta} +\theta_{\nu\beta}\omega_{\mu\alpha}\right),
\label{25}
\end{equation}
\begin{equation}
P^{(0-s)}_{\mu\nu,\alpha\beta} = \frac{1}{3}\theta_{\mu\nu}\theta_{\alpha\beta},
\label{26}
\end{equation}
\begin{equation}
P^{(0-w)}_{\mu\nu,\alpha\beta} = \omega_{\mu\nu}\omega_{\alpha\beta},
\label{27}
\end{equation}
\begin{equation}
P^{(0-sw)}_{\mu\nu,\alpha\beta} = \frac{1}{\sqrt{3}}\theta_{\mu\nu}\omega_{\alpha\beta},
\label{28}
\end{equation}
\begin{equation}
P^{(0-ws)}_{\mu\nu,\alpha\beta} = \frac{1}{\sqrt{3}}\omega_{\mu\nu}\theta_{\alpha\beta},
\label{29}
\end{equation}
are the spin projectors, with $\theta_{\mu\nu} = \eta_{\mu\nu} 
- \omega_{\mu\nu}$ and $\omega_{\mu\nu} = \partial_{\mu}\partial_{\nu}/\bar{\Box}^2$ 
being the transverse and longitudinal projectors.

By inverting the kinetic matrix of $h^{\mu\nu}$ shown in (\ref{23}) and going 
over to momentum space, we obtain the MCG graviton propagator
\begin{eqnarray}
D_{\mu\nu,\alpha\beta} \!\!&=& \!\! -i\,\Bigg\{\frac{2P^{(2)}_{\mu\nu,\alpha\beta}(p)}
{p^{2}\left( p^2 + m^{2}\right)} +\frac{\xi_{1}\left[2P^{(1)}_{\mu\nu,\alpha\beta}(p) 
+ 4P^{(0-w)}_{\mu\nu,\alpha\beta}(p) \right]}{p^{2}\left( p^2 + m^{2}\right)}
\nonumber \\ &&
- \frac{\xi_{2}\left[P^{(0-s)}_{\mu\nu,\alpha
\beta}(p) + \sqrt{3}\left(P^{(0-sw)}_{\mu\nu,\alpha\beta}(p) 
+ 3 P^{(0-ws)}_{\mu\nu,\alpha\beta}(p)\right) 
+ 3P^{(0-w)}_{\mu\nu,\alpha\beta}(p)\right]}{p^{2}\left( p^2 + \xi_{2}m^{2}\right)} 
\Bigg\}. \nonumber \\ &&
\label{30}
\end{eqnarray}
The $p^{-4}$ behavior of all terms of (\ref{30}) at high energies makes 
the theory power-counting renormalizable\footnote{The 
propagator (\ref{30}) reduces to the propagator considered in 
Refs. \cite{Faria6,Faria7} after the imposition of the Feynman gauge 
$\xi_{1} = \xi_{2} = 1$.} \cite{Faria6,Faria7}. Explicitly, the one-loop MCG
divergences are given by\footnote{This result corresponds to the one derived 
in Ref. \cite{Frad} with $\lambda = 0$.} 
\cite{Frad}
\begin{eqnarray}
\Gamma^{(1)}_{\textrm{MCG}} &=&
-\frac{1}{(4\pi)^{2}\varepsilon} \int{d^{4}x} \, \sqrt{-g} 
\Bigg[ \frac{103}{45}E + \frac{797}{60}W - \frac{11}{72}\varphi^{-4}
\left(\varphi^2R-6\varphi\Box\varphi\right)^2  \nonumber \\ && 
- \, \frac{13}{6}\alpha^2\left(\varphi^2R
-6\varphi\Box\varphi\right) + \frac{5}{2}\alpha^4\varphi^4 \Bigg],
\label{31}
\end{eqnarray}
where $\varepsilon$ is the dimensional regularization parameter.

Using the classical background field equation (\ref{6}) in 
(\ref{31})\footnote{In order to (\ref{6}) remain valid even in the presence of 
matter, which will be necessary for the renormalizability of the full theory, 
we will assume that the dilaton field does not couple with matter in Section 
\ref{sec6}.}, and considering the relation (\ref{3}), we obtain the on-shell 
divergent part of the one-loop MCG effective action
\begin{equation}
\Gamma^{(1)\textrm{on-shell}}_{\textrm{MCG}} =
-\frac{1}{(4\pi)^{2}\varepsilon} \int{d^{4}x} \, \sqrt{-g} 
\left[ \frac{797}{120}C^2 - \frac{1567}{360} E    
+ \frac{5}{2}\alpha^4\varphi^4 \right].
\label{32}
\end{equation}
Since the first two terms of (\ref{32}) are of the same type presented 
in the original action (\ref{2}), they are renormalizable. In order to 
renormalize the last term of (\ref{32}), we must add  to (\ref{2}) a quartic 
self-interacting term of the dilaton field $\lambda\int{\sqrt{-g} \varphi^4}$. 
However, the inclusion of this term makes the flat metric no longer a solution 
of the field equations, which invalidates the $S$-matrix formulation. 
Fortunately, this problem is solved if we consider the renormalized value of 
$\lambda$ equal zero so that the self-interacting term is present in the 
renormalized action only to cancel out divergent terms like the last one of 
(\ref{32}). In this way, the theory is one-loop renormalizable. 

Inserting the on-shell MCG effective action up to one-loop into 
the trace of the energy-momentum tensor
\begin{equation}
T = g^{\mu\nu}T_{\mu\nu} = g^{\mu\nu}\frac{2}{\sqrt{-g}}\frac{\delta 
\Gamma_{\textrm{eff}}}{\delta g^{\mu\nu}},
\label{33}
\end{equation}
it can be shown that MCG has the conformal (trace) anomaly
\begin{equation}
T = \frac{1}{(4\pi)^2}\left[\frac{797}{120}C^2 - \frac{1567}{360} E    
+ \frac{5}{2}\alpha^4\varphi^4 \right] 
\neq 0,
\label{34}
\end{equation}
which breaks the conformal symmetry of the theory at the one-loop level. 
A possible consequence of this symmetry breaking is the emergence of 
non-renormalizable $\int{\sqrt{-g}R^2}$, Einstein-Hilbert and cosmological 
constant divergent terms beyond the one-loop level. However,  
by performing the background conformal transformations 
\begin{equation}
\tilde{g}_{\mu\nu} = \left(\varphi/\varphi_{c}\right)^2 g_{\mu\nu}, 
\ \ \ \ \ \ \ \tilde{\varphi} = \varphi_{c},
\label{35}
\end{equation}
we can turn these divergent terms into the same types as the last three terms 
in (\ref{31}). Thus, despite having a conformal anomaly, MCG is 
completely renormalizable.


\subsection{Unitarity}
\label{sec4}


Writing the propagator (\ref{30}) in the form 
\begin{equation}
D_{\mu\nu,\alpha\beta} = -i\frac{2}{m^2} \left[ \frac{1}{p^2} -
\frac{1}{p^2 + m^{2}}\right]P^{(2)}_{\mu\nu,\alpha\beta}(p) 
+ \, \textrm{gauge terms},
\label{36}
\end{equation}
we can see that the renormalizability of the theory is achieved 
at the cost of the emergence of a massive ghost pole at $p^2 = -m^2$ with 
negative residue, in addition to the usual massless graviton pole at $p^2 = 0$ 
with positive residue. In most cases, the presence of a ghost violates the 
unitarity of the theory. However, the fact that the mass of the MCG ghost is 
above the normal threshold of the massless graviton production makes it 
unstable. In this case, the ordinary perturbation theory breaks down and we 
must use a modified perturbation series in which the bare propagator $D(p^2)$ 
is replaced by the dressed propagator 
\cite{Antoniadis}
\begin{equation}
\overline{D}(p^2) = \left[ D^{-1}(p^{2}) - \Pi(p^2) \right]^{-1},
\label{37}
\end{equation}
where $\Pi(p^2)$ is the sum of all one-particle irreducible (1PI) self-energy 
parts. 

We can find the dressed MCG graviton propagator by coupling $N$ fermionic 
fields to the action (\ref{13}), carrying out a $1/N$ expansion, and using 
the Cauchy's integral theorem. The result of such calculation can be write 
in the spectral form \cite{Tomboulis}
\begin{eqnarray}
\overline{D}_{\mu\nu,\alpha\beta} &=& -i\frac{2}{m^2}\left[ \frac{1}{p^2} 
+ \frac{\mathscr{R}}{p^2 - M^{2}} + \frac{\mathscr{R^*}}{p^2 - M^{*2}} 
+ \frac{1}{2\pi}\int_{C}\frac{\rho(a)}{p^2-a}\, da\right]
P^{(2)}_{\mu\nu,\alpha\beta}(p) 
\nonumber \\ && + \, \textrm{gauge terms},
\label{38}
\end{eqnarray}
where $M$ and $M^*$ are the positions of a complex-conjugate pole pair, 
$\mathscr{R}$ and $\mathscr{R}^*$ are the corresponding residues of 
the complex poles, $\rho(a)$ is a spectral function of the 
continuum states and $C$ is an appropriate path in the complex plane.

In the place of the unstable massive ghost pole found in the bare propagator 
(\ref{36}), the dressed propagator (\ref{38}) has a pair of 
complex-conjugate poles in the physical riemannian energy sheet. By using the 
Becchi-Rouet-Stora-Tyutin (BRST) method \cite{Becchi1,Becchi2,Becchi3,Tyutin}, 
and the Ward-Takahashi identities \cite{Ward,Takahashi}, we can find the Nielsen 
identities \cite{Nielsen} for the position of the complex massive 
pole $M$\footnote{The presence of the extra $\Box$ in (\ref{19}) does not affect 
the results of Ref. \cite{Faria8}.} \cite{Faria8}
\begin{equation}
\frac{\partial M^{2}}{\partial\xi_{1}} = 0,
\label{39}
\end{equation}
\begin{equation}
\frac{\partial M^{2}}{\partial\xi_{2}} 
+ C(\left\langle\sigma\right\rangle,\xi_{2})
\frac{\partial M^{2}}{\partial\left\langle\sigma\right\rangle} \neq 0,
\label{40}
\end{equation}  
where $C(\left\langle\sigma\right\rangle,\xi_2)$ is determined 
order by order in the loop expansion of the theory and 
$\left\langle\sigma\right\rangle$ is the vacuum expectation value (VEV) 
of $\sigma$.   

According to (\ref{39}) and (\ref{40}), the position of the complex massive 
pole $M$ is independent of $\xi_1$ but depends on $\xi_2$. Similarly, we can 
show that the same is valid for the position of the complex massive pole 
$M^{*}$. This means that we can move the positions of the complex massive 
poles around by varying $\xi_2$, which causes the excitations represented 
by the complex poles do not contribute to the gauge-invariant absorptive 
part of the $S$-matrix. Thus, the $S$-matrix connects only asymptotic states 
with positive norm, leading to the unitarity of the theory.


\subsection{Symmetry breaking}
\label{sec5}


In order to verify the possibility for spontaneous symmetry breaking in 
the MCG universe, we need to analyze the effective potential 
of the dilaton field, which is given by \cite{Matsuo}
\begin{eqnarray}
V^{(1)}_{\textrm{eff}} 
&=& \sqrt{-g}\left\langle\varphi\right\rangle^{2}R +\frac{\sqrt{-g}}{2(4\pi)^2}
\Bigg\{\left( \log{\frac{\left\langle\varphi\right\rangle^{2}}{\mu^{2}}} 
+ \frac{3}{2} \right)\left(\frac{10}{9}C^2-\frac{985}{72}W\right) \nonumber 
\\ && + \frac{1}{2}\left\langle\varphi\right\rangle^{4} \Bigg[ 5\alpha^{4} 
\left( \log{\frac{\left\langle\varphi\right\rangle^{2}}{\mu^{2}}} 
- \frac{25}{6} \right) +\frac{144}{\mu^4}\left(\frac{10}{9}C^2-\frac{985}{72}
W \right)\Bigg] \nonumber \\ &&  
- \frac{24}{\mu^2}\left\langle\varphi\right\rangle^{2}
\left(\frac{10}{9}C^2-\frac{985}{72}W\right) \Bigg\},
\label{41}
\end{eqnarray}
where $\left\langle\varphi\right\rangle$ is the VEV of 
the dilaton field and $\mu$ is a renormalization mass scale. 

Considering that the metric of the MCG universe is given by\footnote{It is worth 
noting that due to the MCG energy continuity equation $$\frac{d}{dt}
\left[\left(c^{2}\rho + p \right) a^4 \right] = 0,$$ the metric (\ref{42}) 
is valid in all epochs of the MCG universe.} \cite{Faria5}
\begin{equation}
ds^{2} = - c^2dt^{2} + a(t)^2\left( \frac{dr^{2}}{1+r^{2}} 
+ r^{2}d\theta^{2} + r^{2}\sin^{2}\theta d\phi^{2} \right),
\label{42}
\end{equation}
where $a(t) = \sqrt{bt+t^2}$ and $b$ is a positive constant, we find that
\begin{equation}
R = C = 0, \ \ \ \ \ \ \ W < 0.
\label{43}
\end{equation}
In this case, the effective potential (\ref{41}) has a minimum 
$\left\langle\varphi\right\rangle = \varphi_{0}$ away from the origin, which 
means that quantum corrections of the dilaton 
field spontaneously breaks the conformal symmetry via the Coleman-Weinberg 
mechanism \cite{Coleman}. 

After the spontaneous breaking of the conformal symmetry, the MCG action 
(\ref{2}) becomes
\begin{equation}
S_{\textrm{MCG}} = 
\int{d^{4}x} \, \sqrt{-g}\left[ \frac{M_{\textrm{P}}^{2}}{2}R 
- \frac{1}{2\alpha^2}C^2 - \eta E  \right],
\label{44}
\end{equation}
where
\begin{equation}
M_{\textrm{P}}^{2} = 2\varphi_{0}^{2}
\label{45}
\end{equation}
is the reduced Planck mass. It follows from (\ref{45}) that 
the conformal symmetry of the theory is spontaneously broken near the Planck 
scale.


\section{Matter}
\label{sec6}


Here, we follow the GR nomenclature and call ``matter" 
all the fields of nature that are source of gravity, with the exception of the 
gravitational field itself. As in the case of the gravitational action, the 
actions of the matter fields should be conformally invariant in the model 
considered here. In addition, we consider that such actions couple with gravity 
through the interaction of the matter fields with the metric field only and 
not with the dilation field, as stated early in Subsection \ref{sec3}. The 
description of the matter actions conformally coupled with gravity follows 
below.


\subsection{Yang-Mills}
\label{sec7}


The Yang-Mills (YM) theory \cite{Yang} describes the dynamics of $12$ gauge 
fields (spin-$1$ bosons), namely, $1$ photon $A_{\mu}$, which mediates the 
electromagnetic interaction, $3$ weak bosons $W^{\pm}_{\mu}$ and $Z^{0}_{\mu}$, 
which mediate the charged and neutral current weak interactions, and $8$ 
gluons $G^{a}_{\mu}$ ($a = 1, 2, \ldots, 8$), which mediate the strong 
interactions. Since the standard YM action in flat spacetime is already 
conformally invariant, its conformal coupling with gravity is performed 
only through the minimal coupling $\eta_{\mu\nu} \rightarrow g_{\mu\nu}$ 
and $d^{4}x \rightarrow d^{4}x\sqrt{-g}$.

The YM action conformally coupled with gravity is given by 
\begin{equation}
S_{\textrm{CYM}} =  -\frac{1}{4}\int{d^{4}x} \, \sqrt{-g}\left[g^{\mu\alpha}
g^{\nu\beta} \left( B_{\mu\nu}B_{\alpha\beta} 
+ W^{i}_{\mu\nu}W^{i}_{\alpha\beta} + G^{a}_{\mu\nu}G^{a}_{\alpha\beta} 
\right) \right],
\label{46}
\end{equation}
where
\begin{equation}
B_{\mu\nu} = \partial_{\mu}B_{\nu} - \partial_{\nu}B_{\mu}
\label{47}
\end{equation}
is the field strength of the $B_{\mu}$ gauge field that corresponds to the 
phase transformations $U(1)_{Y}$ symmetry group,
\begin{equation}
W^{i}_{\mu\nu} = \partial_{\mu}W^{i}_{\nu} - \partial_{\nu}W^{i}_{\mu}
- \eta g\epsilon^{ijk}W^{j}_{\mu}W^{k}_{\nu} 
\label{48}
\end{equation}
is the field strength of the $W^{i}_{\mu}$ ($i= 1, 2, 3$) gauge fields 
that correspond to the chiral $SU(2)_{L}$ symmetry group, and
\begin{equation}
G^{a}_{\mu\nu} = \partial_{\mu}G^{a}_{\nu} - \partial_{\nu}G^{a}_{\mu}
- \eta_{s} g_{s}f^{abc}G^{b}_{\mu}G^{c}_{\nu}
\label{49}
\end{equation}
is the field strength of the $G^{a}_{\mu}$ gauge fields that correspond 
to the color $SU(3)_{C}$ symmetry group. 

In (\ref{48}) and (\ref{49}), we have that $\eta = \pm 1$, $g$ is the 
coupling constant associated with the $SU(2)_{L}$ group, $\epsilon^{ijk}$ 
is the Levi-Civita symbol, $\eta_{s} = \pm 1$, $g_{s}$ is the coupling 
constant associated with the $SU(3)_{C}$ group, and $f^{abc}$ are the 
structure constants of the $SU(3)_{C}$ group, satisfying $[T^{a},T^{b}] 
= i f^{abc}T^{c}$, with $T^{a}$ being the generators of the $SU(3)_{C}$ 
group, which can be represented by the Gell-Mann matrices $\lambda^{a}$ 
according to $T^{a} = (1/2)\lambda^{a}$. Note that the generic signs 
in the $\eta$ parameters are necessary to specify the different notations 
found in literature.

\subsection{Higgs}
\label{sec8}


The Higgs field was introduced in the SM  for the mass terms of the $W$ and 
$Z$ gauge bosons, and fermions, to become $SU(2)_{L} \times U(1)_{Y}$ invariant 
\cite{Brout,Higgs}. However, the explicit mass term of the Higgs boson breaks 
the conformal symmetry of the theory. This problem can be fixed if we consider 
that the Higgs mass is generated by the symmetry breaking of an additional 
scalar field. 

The Higgs action conformally coupled with gravity that we 
consider here is
\begin{eqnarray}
S_{\textrm{CH}} &=& - \int{d^{4}x} \, \sqrt{-g}\bigg[ 
 g^{\mu\nu}\left(D_{\mu} \mathcal{H}\right)^{\dagger} D_{\nu} \mathcal{H} 
 + \frac{1}{2}g^{\mu\nu}\partial_{\mu}S\partial_{\nu}S 
+ \frac{1}{2}g^{\mu\nu}\partial_{\mu}\phi \partial_{\nu}\phi \bigg] 
\nonumber \\ &&-\, \int{d^{4}x} \, V(\mathcal{H},S,\phi),
\label{50}
\end{eqnarray}
where
\begin{equation}
\mathcal{H} = \begin{pmatrix}H^{+}\\H^{0}\end{pmatrix} 
\label{51}
\end{equation}
is the Higgs doublet field with $2$ complex scalar fields $H^{+}$ and 
$H^{0}$, $S$ and $\phi$ are real scalar singlet fields\footnote{The addition 
of two scalars instead of just one will be justified in Section \ref{sec11}.}, 
\begin{equation}
D_{\mu} \mathcal{H} = \left(\partial_{\mu} + i\eta g T^{i} W^{i}_{\mu} 
+ i\eta' g' \eta_{Y}Y B_{\mu} \right) \mathcal{H}
\label{52}
\end{equation}
is the $SU(2)_{L} \times U(1)_{Y}$ gauge covariant derivative for the Higgs 
field, and
\begin{eqnarray}
V(\mathcal{H},S,\phi) &=& \sqrt{-g} \bigg[\frac{1}{6}\left(\mathcal{H}^{\dagger}
\mathcal{H}\right) R + \frac{1}{12}S^{2}R + \frac{1}{12}\phi^{2}R 
+\frac{\lambda_{\mathcal{H}}}{4}\left(\mathcal{H}^{\dagger}\mathcal{H}\right)^2 
+\frac{\lambda_{S}}{4}S^{4} \nonumber \\ && + \frac{\lambda_{\phi}}{4}\phi^{4} 
+ \frac{\lambda_{\mathcal{H}S}}{2}S^{2}\left(\mathcal{H}^{\dagger}\mathcal{H}
\right) +\frac{\lambda_{\mathcal{H}\phi}}{2}\phi^{2}\left(\mathcal{H}^{\dagger}
\mathcal{H}\right) + \frac{\lambda_{S\phi}}{2}S^{2}\phi^{2}\bigg] 
\label{53}
\end{eqnarray}
is the scalar potential, with $\lambda_{\mathcal{H}}$, 
$\lambda_{S}$, $\lambda_{\phi}$, $\lambda_{\mathcal{H}S}$, 
$\lambda_{\mathcal{H}\phi}$ and $\lambda_{S\phi}$ being  dimensionless 
coupling constants.

We have in (\ref{52}) that $T^{i}$ are the generators of the 
$SU(2)_{L}$ group, satisfying $[T^{i},T^{j}] = i \epsilon^{ijk}T^{k}$, 
$\eta' = \pm 1$, $g'$ is the coupling constant associated with the $U(1)_{Y}$
group, $\eta_{Y} = \pm 1$, and the hypercharge $Y$ is the generator of the
$U(1)_{Y}$ group, which is related to the electron charge $Q$ and the $T^{3}$ 
generator of the $SU(2)_{L}$ group by $Q = T^{3} + \eta_{Y}Y$. We can represent 
the generators of the $SU(2)_{L}$ group in the form $T^{i} = (i/2)\sigma^{i}$, 
where $\sigma^{i}$ are the Pauli matrices.


\subsection{Fermions}
\label{sec9}


The fermions are composed by $6$ leptons $e_{M}$ and $\nu_{M}$, 
and $6$ quarks $u^{C}_{M}$ and $d^{C}_{M}$, where $M = 1, 2, 3$ is the 
generation index such that $e_{M}$ is the electron, the muon or the tau, 
$\nu_{M}$ is the corresponding neutrino,  $u^{C}_{M}$ is either the up, 
charm or top quark, and $d^{C}_{M}$ is the down, strange or bottom quark, 
with $C = 1, 2, 3$ corresponding to the three types of $SU(3)$ color
$R$, $G$, $B$. 

By conformally coupling the SM fermion action with gravity, we find
\begin{equation}
S_{\textrm{CF}} = - \int{d^{4}x} \, \sqrt{-g}\left(
 \sum_{\psi_{L}}  i\bar{\psi}_{L}\slashed{D}\psi_{L} 
+ \sum_{\psi_{R}} i\bar{\psi}_{R}\slashed{D}\psi_{R} \right),
\label{54}
\end{equation}
where summation over the generation and color indices is implied,
$\bar{\psi} = \psi^{\dagger}\gamma^{0}$ is the 
adjoint fermion field,
\begin{equation}
\slashed{D}\psi_{L} = e^{\mu}\,\!\!_{a}\gamma^{a} \left(\partial_{\mu} 
+ i\eta g T^{i} W^{i}_{\mu} + i\eta' g' \eta_{Y}Y B_{\mu}
- \frac{i}{4}\omega_{\mu ab}\left[\gamma^{a},
\gamma^{b}\right]\right)\psi_{L}
\label{55}
\end{equation}
is the full gauge covariant derivative for the left-handed fermions
\begin{equation}
\psi_{L} = L, Q = 
\begin{pmatrix}\nu_{L} \\ e_{L}\end{pmatrix} \, , \,
\begin{pmatrix}u_{L} \\ d_{L}\end{pmatrix},
\label{56}
\end{equation}
and
\begin{equation}
\slashed{D}\psi_{R} = e^{\mu}\,\!\!_{a}\gamma^{a} \left(\partial_{\mu} 
+ i\eta' g' \eta_{Y}Y B_{\mu} 
- \frac{i}{4}\omega_{\mu ab}\left[\gamma^{a},\gamma^{b}\right]
\right)\psi_{R}
\label{57}
\end{equation}
is the full gauge covariant derivative for the right-handed fermions
\begin{equation}
\psi_{R} =  \nu_{R}, e_{R}, u_{R}, d_{R}.
\label{58}
\end{equation}

In (\ref{55}) and (\ref{57}), we have that $e_{\mu}\,\!\!^{a}$ are the tetrad 
fields, $\gamma^{a}$ are the Dirac matrices, which satisfy the anticommutation 
relation $\{ \gamma^{a}, \gamma^{b}\} = 2\eta^{ab}$, and 
$\omega_{\mu}\,\!\!^{a}\,\!_{b} = e_{\lambda}\,\!\!^{a}e^{\nu}\,\!\!_{b}
\Gamma^{\lambda}\,\!\!_{\mu\nu} 
- e^{\lambda}\,\!\!_{b}\partial_{\mu}e_{\lambda}\,\!\!^{a}$ is the spin 
connection.


\subsection{Yukawa}
\label{sec10}


The SM Yukawa action describes the interaction of the fermions with the 
Higgs field. These interactions are responsible for generating the masses 
of the fermions, with the exception of the neutrinos,  when quantum 
corrections of the Higgs field spontaneously breaks the $SU(2)_{L} 
\times U(1)_{Y}$ (electroweak) symmetry. Including an extra interaction of 
the Higgs field with the neutrinos, the latter acquire similar masses to the 
other particles in the SM instead of the small masses found in nature. One way 
to solve this problem is through the seesaw mechanism \cite{Gell}, which 
assumes the SM interactions of the fermions with the Higgs field and 
introduces a Majorana mass term for the right-handed neutrinos. However, 
the Majorana mass term breaks the conformal symmetry of the theory. Therefore, 
in order to make the theory conformally invariant, we can consider that the 
neutrino masses are generated by the symmetry breaking of one of the extra 
scalars, from which we choose $S$.

The resulting conformally invariant Yukawa action coupled with gravity reads
\begin{eqnarray}
S_{\textrm{CY}} &=& - \int{d^{4}x} \, \sqrt{-g}\bigg( 
\bar{L}Y^{e}\mathcal{H}e_{R} 
+ \bar{Q}Y^{u} \tilde{\mathcal{H}}u_{R} 
+ \bar{Q}Y^{d}\mathcal{H}d_{R}\nonumber \\ &&   
+ \, \bar{L}Y^{\nu} \tilde{\mathcal{H}}\nu_{R}
+ \nu^{T}_{R}Y^{M}\mathcal{C}S\nu_{R} + \textrm{h.c.} \bigg),
\label{59}
\end{eqnarray}
where summation over the generation and color indices is assumed, $Y^{e}$, 
$Y^{u}$ and $Y^{d}$ are the usual Yukawa matrices of the SM, $Y^{\nu}$ is a 
complex matrix that mediates the coupling of the SM fields to the 
right-handed neutrinos, $Y^{M}$ is a real diagonal matrix that describes 
the interactions of the right-handed neutrinos with the scalon 
and $\tilde{\mathcal{H}} = i\sigma_{2}\mathcal{H}^{\dagger}$.


\section{Full theory}
\label{sec11}


Finally, putting together all the pieces of the previous sections, 
we arrive at the CTOE action
\begin{equation}
S_{\textrm{CTOE}} = S_{\textrm{MCG}} + S_{\textrm{CYM}} 
+ S_{\textrm{CH}} + S_{\textrm{CF}} + S_{\textrm{CY}},
\label{60}
\end{equation}
where $S_{\textrm{MCG}}$, $S_{\textrm{CYM}}$, $S_{\textrm{CH}}$, 
$S_{\textrm{CF}}$ and $S_{\textrm{CY}}$ are given by (\ref{2}), (\ref{46}), 
(\ref{50}), (\ref{54}) and (\ref{59}), respectively. Note that we not 
explicitly consider possible gauge-fixing terms as well as the compensating 
Faddeev-Popov ghost terms in (\ref{60}).

It can be shown that the contribution of the matter fields to the on-shell 
divergences of the one-loop CTOE effective action are of the same types 
presented in (\ref{60}) \cite{Buch3}, which means that the full theory 
is one-loop renormalizable. 
The non-renormalizable divergences that may arise beyond the one-loop level 
due to the conformal anomaly produced by the contribution of the matter 
fields can be eliminated by performing the conformal transformations 
(\ref{35}) and using the field equation (\ref{6}), which makes CTOE 
completely renormalizable. Additionally, since there are no ghosts in 
the matter part of (\ref{60}), the theory is also unitary.

The investigation of the spontaneous symmetry breaking in the CTOE matter part 
requires the minimization of the one-loop matter effective potential of the 
theory, which is a quite cumbersome thing to do due to the existence of three 
interacting scalar fields coupled with gravity in (\ref{53}). However, in flat 
spacetime, the potential (\ref{53}) reduces to
\begin{equation}
V(h,S,\phi) = \frac{\lambda_{h}}{4}h^4 
+\frac{\lambda_{S}}{4}S^{4}  + \frac{\lambda_{\phi}}{4}\phi^{4} 
+ \frac{\lambda_{hS}}{2}h^2S^{2} 
+\, \frac{\lambda_{h\phi}}{2}h^2\phi^{2} 
+ \frac{\lambda_{S\phi}}{2}S^{2}\phi^{2},
\label{61}
\end{equation}
where $h^{2} = \mathcal{H}^{\dagger}\mathcal{H}$. Using the Gildener-Weinberg 
(GW) analytical approximation method \cite{GW}, it was 
shown in Ref. \cite{Helm} that the potential (\ref{61}) leads to the correct
phenomenology at low energies, while keeping the system stable\footnote{The 
effects of the top quark Yukawa coupling on the stability of the renormalization 
group (RG) running still need to be studied.} and free of Landau poles up to 
the Planck scale, if we assume that $\phi$ does not develop a VEV and $S$ is a 
pseudo-Goldstone boson (PGB) that acquires a finite VEV during the electroweak 
symmetry breaking. 

After the spontaneous breaking of the electroweak symmetry, the gauge bosons 
and fermions, with the exception of the right-handed neutrinos, acquire masses 
as in the standard model. In addition, the interactions of the Higgs field and 
the right-handed neutrinos with $S$, allow that the observed masses of these 
fields be generated after the spontaneous breaking of the electroweak symmetry.
Since all scales of CTOE are generated 
by spontaneous breaking of symmetry, the theory is free of the 
hierarchy problem.


\section{Final remarks}
\label{sec12}


We have constructed here a renormalizable and unitary 
theory of everything by conformally coupling an extended SM with gravity. In 
the gravitational part of the theory there is a scalar field $\varphi$ called 
dilation whose quantum corrections spontaneously breaks the conformal symmetry 
near the Planck scale. In addition, the electroweak symmetry is spontaneously 
broken by quantum corrections of an extra scalar field $S$ introduced in the 
matter part of the theory, which also contains a second extra scalar $\phi$ 
with zero vacuum expectation value. 

Besides being at energies far beyond the reach of current particle 
accelerators, the dilaton field has no degrees of freedom and thus it 
is not detectable. On the other hand, the preferable energy of $S$ is of 
the order of a few GeV, which leads to additional Higgs decays that can be 
tested in future LHC runs. Since $\phi$, whose energy must lie between $300$ 
GeV and $370$ GeV, does not have a vacuum expectation value, it may be a
good candidate for dark matter, which we intend to study further in future works.

\section*{Acknowledgments}
\label{sec13}

The author is grateful to I. L. Shapiro and A. D. Pereira for very helpful 
discussions. In additon, the author would like to thank S. D. Odintsov, 
E. Elizalde, A. Salvio and A. Strumia for useful comments.



\end{document}